\documentclass[apj]{emulateapj}
\usepackage{color,amsmath,amssymb,latexsym,txfonts}
\usepackage{graphicx,subfigure,threeparttable}
\def\jcap{J.\ Cosmol.\ Astropart.\ Phys.\ }
\def\nar{New\ Astron.\ Rev.\ }

\begin{document}

\title{Testing the millisecond pulsar scenario of the Galactic center
gamma-ray excess with very high energy gamma-rays}

\author{Qiang Yuan$^{1,2}$ and Kunihito Ioka$^{2,3}$}

\affil{$^1$Department of Astronomy, University of Massachusetts,
Amherst, MA 01002, USA\\
$^2$Theory Center, Institute of Particle and Nuclear Studies,
KEK, Tsukuba 305-0801, Japan\\
$^3$Department of Particle and Nuclear Physics, the Graduate 
University for Advanced Studies (Sokendai), Tsukuba 305-0801, Japan}

\begin{abstract}

The recent analyses of the Fermi Large Area Telescope data show an 
extended GeV $\gamma$-ray excess on top of the expected diffuse background 
in the Galactic center region, which can be explained with annihilating 
dark matter or a population of millisecond pulsars (MSPs). We propose to 
observe the very high energy $\gamma$-rays for distinguishing the MSP 
scenario from the dark matter scenario. The GeV $\gamma$-ray MSPs should 
release most energy to the relativistic $e^{\pm}$ wind, which will 
diffuse in the Galaxy and radiate TeV $\gamma$-rays through inverse 
Compton scattering and bremsstrahlung processes. By calculating the 
spectrum and spatial distribution, we show that such emission is 
detectable with the next generation very high energy $\gamma$-ray 
observatory, the Cherenkov Telescope Array (CTA), under reasonable 
model parameters. It is essential to search for the multi-wavelength 
counterparts to the GeV $\gamma$-ray excess for solving this mystery 
in the high energy universe.

\end{abstract}

\section{Introduction}

The observations of high energy $\gamma$-rays from the Galactic center 
(GC) have revealed very interesting features 
at different scales. The atmospheric imaging Cherenkov telescope array,
High Energy Stereoscopic System (HESS), has discovered a point-like 
source\footnote{See \citet{2004ApJ...608L..97K} for an earlier 
potential detection of it with significance $\sim3.7\sigma$ by Whipple.} 
which may be originated from the central supermassive black hole
\citep{2004A&A...425L..13A} as well as an extended source from the
ridge which could be explained through the interaction of a fresh
cosmic ray (CR) source with the interstellar medium (ISM)
\citep{2006Natur.439..695A}. The analyses of the data from the space-borne 
$\gamma$-ray detector, Fermi Large Area Telescope (Fermi-LAT), have also
revealed the corresponding point-like \citep{2011ApJ...726...60C} and ridge 
emission \citep{2013ApJ...762...33Y,2014PhRvD..89f3515M}. Moreover, an even 
more extended, circular $\gamma$-ray excess on top of the central point 
and ridge emission has been identified in the Fermi-LAT data by quite a few 
groups \citep{2009arXiv0910.2998G,2009arXiv0912.3828V,2011PhLB..697..412H,
2011PhLB..705..165B,2012PhRvD..86h3511A,2013PhRvD..88h3521G,
2013PDU.....2..118H,2013arXiv1307.6862H,2014PhRvD..90b3526A,
2014arXiv1402.6703D,2014arXiv1406.6948Z,2014arXiv1409.0042C,
2014Fermi-GC}. Finally, the bubble structure on the Galactic scale 
has also been shown in the Fermi-LAT data \citep{2010ApJ...717..825D,
2010ApJ...724.1044S,2012ApJ...753...61S,2014A&A...567A..19Y,
2014ApJ...793...64A}. Those multi-scale, wide-band features have very
interesting implications of the GC activity and stellar history
\citep{2011MNRAS.413..763C,2012MNRAS.423.3512C}.

The circularly symmetric $\gamma$-ray excess is of special interest 
since it may connect with the emission from dark matter (DM) annihilation 
in the GC \citep{2009arXiv0910.2998G,2011PhLB..697..412H,
2012PhRvD..86h3511A,2013PhRvD..88h3521G,2014PhRvD..89a5023H,
2014PhRvD..90b3526A,2014arXiv1402.6703D}. The de-projected
spatial distribution of the $\gamma$-ray source follows approximately
$\theta^{-2.4}$,  with $\theta$ the angle away from the GC, which seems to be 
consistent with the DM annihilation with a generalized Navarro-Frenk-White 
\citep[gNFW,][]{1997ApJ...490..493N,1996MNRAS.278..488Z} density profile.
The energy spectrum peaks at several GeV, and can be well fitted with the
scenario of DM annihilation into a pair of quarks or $\tau$ leptons with
mass of DM particles $10$s GeV \citep{2011PhLB..697..412H,2012PhRvD..86h3511A,
2013PhRvD..88h3521G}. Nevertheless, the astrophysical scenarios such as
a population of unresolved millisecond pulsars\footnote{See, however,
\citet{2013PhRvD..88h3009H} and \citet{2014arXiv1407.5625C}.} \citep[MSPs,]
[]{2005MNRAS.358..263W,2011JCAP...03..010A,2013MNRAS.436.2461M,
2014JHEAp...3....1Y,2014ApJ...796...14C}, as well as the CR interactions 
in the GC may also account for this $\gamma$-ray excess 
\citep{2014PhRvD..90b3015C,2014JCAP...10..052P}. On the other hand, other 
observations of CRs and radio emission seem to have tension with the DM 
annihilation scenario, although they still suffer from uncertainties 
\citep{2014arXiv1406.6027B,2014arXiv1407.2173C,2014arXiv1410.1527H}. 
The origin of the $\gamma$-ray excess is still in debate, and we may need 
future, multi-wavelength observations to test different models.

The MSP scenario is favored for several reasons. The spectrum of the
GeV $\gamma$-ray excess is very similar to those observed for MSPs 
\citep{2013ApJS..208...17A} and for some of the globular clusters 
\citep{2010A&A...524A..75A}. The MSPs are old enough to extend to 
relatively large scale which can possibly explain the large extension 
of the GeV $\gamma$-ray excess \citep{2014arXiv1402.6703D}.
The observed distribution of the Galactic low mass X-ray binaries
\citep{2008A&A...491..209R}, the progenitors of MSPs, just follows
the $\theta^{-2.4}$ profile as observed for the $\gamma$-ray signal 
\citep{2014JHEAp...3....1Y}. If there is a population of unresolved 
$\gamma$-ray MSPs in the Galactic bulge to contribute to the 
$\gamma$-ray excess, one would naturally expect that
a portion (probably most) of the spin-down energy should be carried by 
the relativistic $e^{\pm}$ winds from those pulsars. Such wind $e^{\pm}$
may further be accelerated up to very high energies (VHE; TeV energies) 
in the shocks produced by the pulsar winds \citep{2007MNRAS.377..920B,
2013MNRAS.435L..14B}. Then these leptons will diffuse into the 
ISM in the Galaxy, and radiate through inverse Compton scattering (ICS), 
bremsstrahlung and synchrotron emission. Since the $e^{\pm}$ wind 
generally dominates the energy budget over the GeV $\gamma$-rays from 
pulsars, we should investigate such emission from $e^{\pm}$. The future 
observations of the VHE $\gamma$-rays from the GC region by e.g., 
the Cherenkov Telescope Array \citep[CTA,][]{2013APh....43....3A}, could 
test the MSP scenario of the Fermi GeV $\gamma$-ray excess, and may 
distinguish it from the DM scenario \citep{2014PhRvD..90d3508L,
2014arXiv1410.6168A}.

In this work we explore the VHE $\gamma$-ray emission associated with the 
relativistic $e^{\pm}$ winds from the GC MSPs, using the Fermi $\gamma$-ray 
excess as a normalization of the MSP population. In Sec. 2 we briefly 
describe the $e^{\pm}$ production model of MSPs. The major results of the
VHE $\gamma$-ray emission from the MSP $e^{\pm}$ and the detectability 
with CTA will be given in Sec. 3. We conclude and discuss the results 
in Sec. 4.

\section{$e^{\pm}$ from MSPs}

It has been well established that the rotation-powered pulsars release
most of their rotational energy as a relativistic wind of magnetized
$e^{\pm}$ plasma from the magnetosphere \citep{1974MNRAS.167....1R}.
The interaction of such a wind with the ISM will then form a shock, which 
can also accelerate the $e^{\pm}$ to VHE. The synchrotron or ICS 
emission from these $e^{\pm}$ form the so-called pulsar wind nebula (PWN).

Assuming that the pulsar spin-down power is predominately carried away by the
kinetic energy of the wind particles, one has \citep{1984ApJ...283..694K,
1984ApJ...283..710K}
\begin{equation}
L_{\rm sd}=\gamma_w\dot{N}_{\rm GJ}m_ec^2\kappa(1+\sigma)/f_{e^{\pm}},
\end{equation}
where $L_{\rm sd}$ is the spin-down power of the pulsar, $\gamma_w$ is
the Lorentz factor of the wind particles, $\dot{N}_{\rm GJ}$ is the
Goldreich-Julian current \citep{1969ApJ...157..869G}, $m_e$ is the electron
mass, $c$ is the speed of light, $\kappa$ is the $e^{\pm}$ pair multiplicity,
$\sigma$ is the magnetization parameter (the ratio of the Poynting flux
to the kinetic energy flux), and $f_{e^{\pm}}$ is the relative energy 
fraction carried by $e^{\pm}$ particles. For typical parameters of MSPs, 
the Lorentz factor of the wind particles can be derived as 
\citep{2006ApJ...641..427C,2010ApJ...723.1219C}
\begin{equation}
\gamma_w=4\times10^5(f_{e^{\pm}}/\kappa_3)L_{34}^{1/2},
\end{equation}
where $L_{34}$ is the spin-down power in unit of $10^{34}$ erg s$^{-1}$,
and $\kappa_3=\kappa/10^3$. The spin-down power carried away by the 
$e^{\pm}$ pairs is simply $L_{e^{\pm}}=f_{e^{\pm}}L_{\rm sd}$. 
Typically we will have $f_{e^{\pm}}\simeq 1$. The energy distribution
of the wind $e^{\pm}$ could be relativistic Maxwellian 
\citep{2012A&A...547A.114A}. For simplicity we adopt monochromatic 
injection of the wind $e^{\pm}$ for the case without considering the
possible acceleration or cooling in the PWN (see below).


The pulsar winds may further be accelerated in the shock region if exists
\citep{2007MNRAS.377..920B,2013MNRAS.435L..14B}. In the GC stellar 
cluster, the maximum achievable energy of the $e^{\pm}$ can be as high as 
$50$ TeV \citep{2013MNRAS.435L..14B}. The fraction of $e^{\pm}$ which are 
accelerated, and the final spectrum of the accelerated $e^{\pm}$ are not 
clear. We may simply assume a power-law spectrum with an exponential 
cutoff at the maximum energy \citep{2013MNRAS.435L..14B}. The total 
power of the accelerated $e^{\pm}$ can be absorbed in the efficiency 
$f_{e^{\pm}}$.

Finally, the $e^{\pm}$ may radiate and lose energy inside the PWN. 
For the young pulsar (e.g., Crab), most of the spin-down energy is 
released through synchrotron radiation in the nebula. However, 
it may not be the case for the MSPs since they are generally old and
their PWNe are so extended that the magnetic field is weak
\citep{2011PhRvD..83b3002K,2012MNRAS.421.3543K}. Therefore we assume
a fraction of the $e^{\pm}$ can escape away from the PWN into the ISM
of the Galaxy, and absorb this fraction in the efficiency $f_{e^{\pm}}$.

As for the spatial distribution of the $e^{\pm}$ injection,
we adopt the result inferred from the $\gamma$-ray excess, i.e., a 
spherical power-law distribution $r^{-2.4}$ which extends to $\sim1.5$ 
kpc ($\sim10^{\circ}$) \citep{2011PhLB..697..412H,2012PhRvD..86h3511A,
2013PhRvD..88h3521G,2014arXiv1402.6703D}. The $\gamma$-ray emission 
($e^{\pm}$ source) may extend to even large scales 
\citep{2014arXiv1402.6703D,2014arXiv1406.6948Z}. However, it will not 
significantly affect the results of the current
study which focuses on the GC region.

In summary, we will discuss either monochromatic or power-law injection 
of $e^{\pm}$ in the extended region around the GC, which represent 
the scenario without or with acceleration in the pulsar wind shocks, 
respectively. The canonical model parameters of the $e^{\pm}$ are 
tabulated in Table \ref{table:model}. Furthermore, we will assume that 
the typical $\gamma$-ray efficiency with respect to the spin-down power is 
$0.1$ \citep{2013ApJS..208...17A}. Thus we have $L_{e^{\pm}}=f_{e^{\pm}} 
L_{\rm sd}=10f_{e^{\pm}} L_{\gamma}$.

\begin{table}[!htb]
\centering
\caption{Injection $e^{\pm}$ parameters: injection energy $E_{\rm inj}$
for the monochromatic case, spectral index $\alpha$ and cutoff energy 
$E_{\rm max}$ for the power-law case, and the $e^{\pm}$ energy fraction of 
the spindown power $f_{e^{\pm}}$.}
\begin{tabular}{ccccc}
\hline \hline
Spectrum & $E_{\rm inj}$ & $\alpha$ & $E_{\rm max}$ & $f_{e^{\pm}}$ \\
 & (GeV) & & (GeV) & \\
\hline
$\delta(E-E_{\rm inj})$ & $200$ & ... & ... & 0.9 or 0.1 \\
$\delta(E-E_{\rm inj})$ & $20$ & ... & ... & 0.9 or 0.1 \\
$E^{-\alpha}\exp(-E/E_{\rm max})$ & ... & $2.0$ & $5\times10^4$ & 0.9 or 0.1 \\
\hline
\hline
\end{tabular}
\label{table:model}
\end{table}

\section{VHE $\gamma$-ray emission of GC MSPs}

Given the above source injection of relativistic $e^{\pm}$, we use 
GALPROP\footnote{http://galprop.stanford.edu/} numerical tool 
\citep{1998ApJ...509..212S,1998ApJ...493..694M} to calculate the
propagation of the $e^{\pm}$ as well as the secondary $\gamma$-ray 
emission through ICS and bremsstrahlung. 
We use the default magnetic field in GALPROP, which is 
about 5 $\mu$G in the GC. Thus the synchrotron energy loss of the 
$e^{\pm}$ is less important compared with the ICS energy loss given the 
very strong optical radiation in the GC \citep[$\sim10$ eV cm$^{-3}$;][]
{2006ApJ...640L.155M}. The diffusion with reacceleration 
configuration of the propagation model is adopted. Three groups of the 
propagation parameters are adopted with the characteristic
height of the propagation halo, $z_h$, varying from 2 to 10 kpc 
\citep{2012ApJ...761...91A}. Such a range will largely cover the 
uncertainties of the propagation parameters \citep{2011ApJ...729..106T,
2014arXiv1410.0171J}. The propagation parameters are given in Table
\ref{table:prop}.

\begin{table}[!htb]
\centering
\caption{Cosmic ray propagation parameters. Columns from left to right
are the diffusion coefficient $D_0$ at the reference rigidity $R=4$ GV, 
height of the propagation halo $z_h$, Alfven speed $v_A$ which characterizes 
the reacceleration, and power-law index $\delta$ of the rigidity dependence
of the diffusion coefficient.}
\begin{tabular}{ccccc}
\hline \hline
 & $D_0$ & $z_h$ & $v_A$ & $\delta$ \\
 & ($10^{28}$cm$^2$\,s$^{-1}$) & (kpc) & (km s$^{-1}$) & \\
\hline
1 & $2.7$ & $2$ & $35.0$ & $0.33$ \\
2 & $5.3$ & $4$ & $33.5$ & $0.33$ \\
3 & $9.4$ & $10$ & $28.6$ & $0.33$ \\
\hline
\hline
\end{tabular}
\label{table:prop}
\end{table}

\begin{figure*}[!htb]
\centering
\includegraphics[width=\columnwidth]{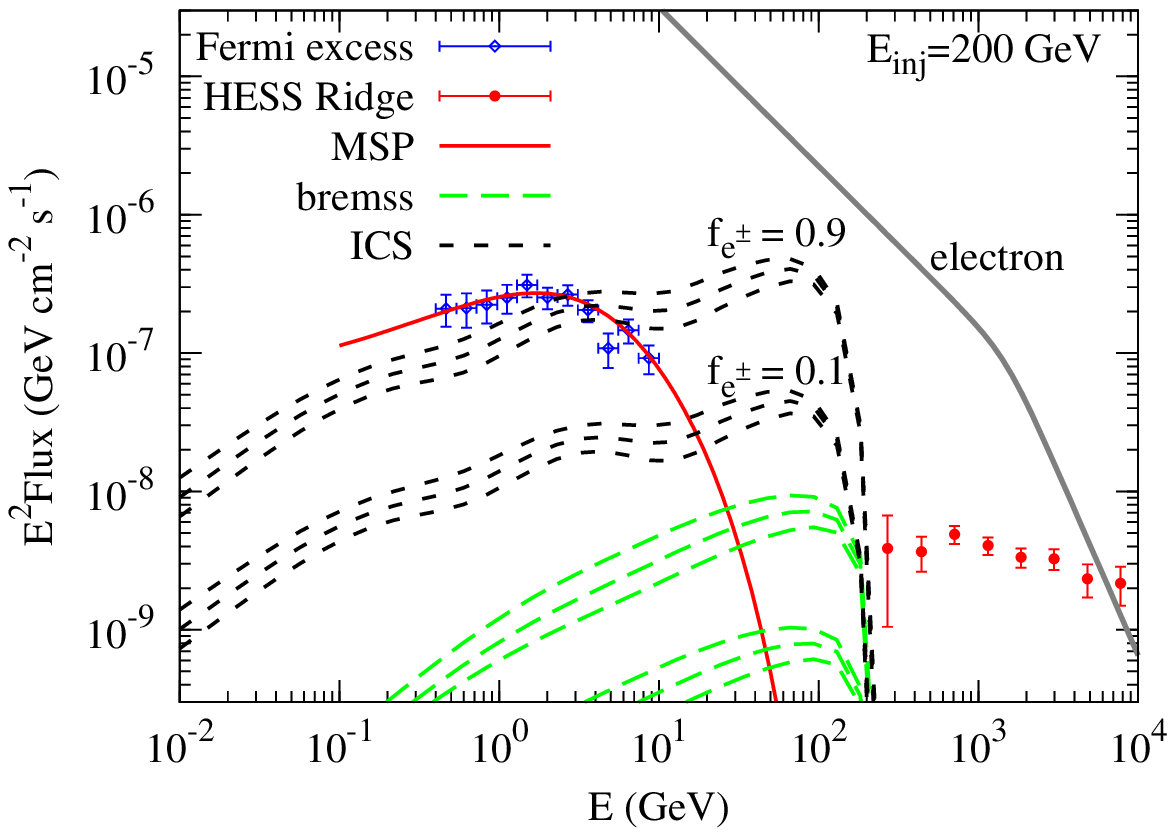}
\includegraphics[width=\columnwidth]{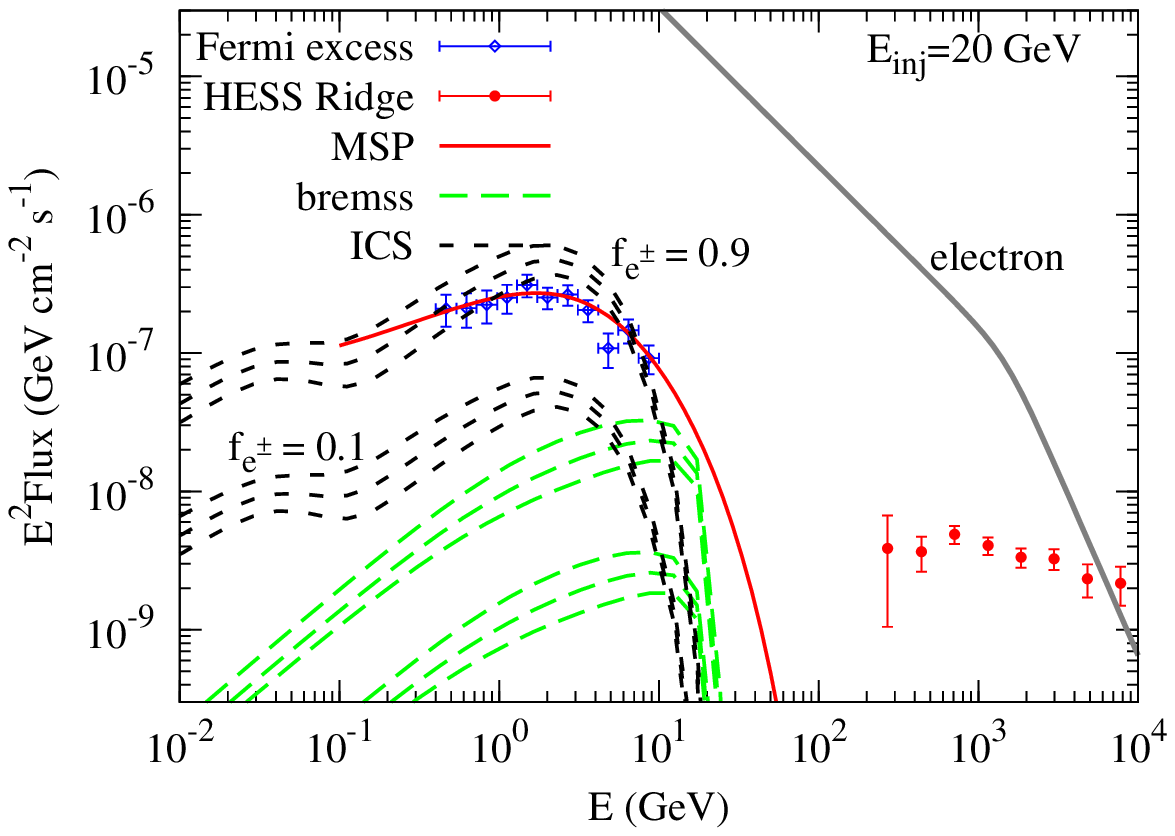}
\caption{Expected ICS (short dashed) and bremsstrahlung (long dashed)  
emission from the wind $e^{\pm}$ of the MSPs in the bulge, compared with
the direct $\gamma$-ray emission from MSPs (solid) and the Fermi-LAT
data of the excess \citep{2013PhRvD..88h3521G}. The three lines of the
ICS and bremsstrahlung groups represent three different settings of 
propagation parameters in Table \ref{table:prop}, with $z_h$ increasing 
from top to bottom. The $\gamma$-ray emission in $7^{\circ}\times 
7^{\circ}$ region around the GC is integrated. The left panel is for 
the monochromatic injection of $e^{\pm}$ with $E_{\rm inj}=200$ GeV, and
the right panel is for $E_{\rm inj}=20$ GeV. As a comparison, the gray
line shows the fluxes of the CR $e^{\pm}$ integrated in the same sky
region according to the AMS-02 \citep{2014PhRvL.113l1102A} and HESS 
\citep{2008PhRvL.101z1104A} data. Also shown the HESS observations 
of the diffuse emission from the GC ridge integrated in the region 
$|l|<0^{\circ}.8$, $|b|<0^{\circ}.3$ \citep{2006Natur.439..695A}.
}
\label{fig:sed_mono}
\end{figure*}

\begin{figure*}[!htb]
\centering
\includegraphics[width=\columnwidth]{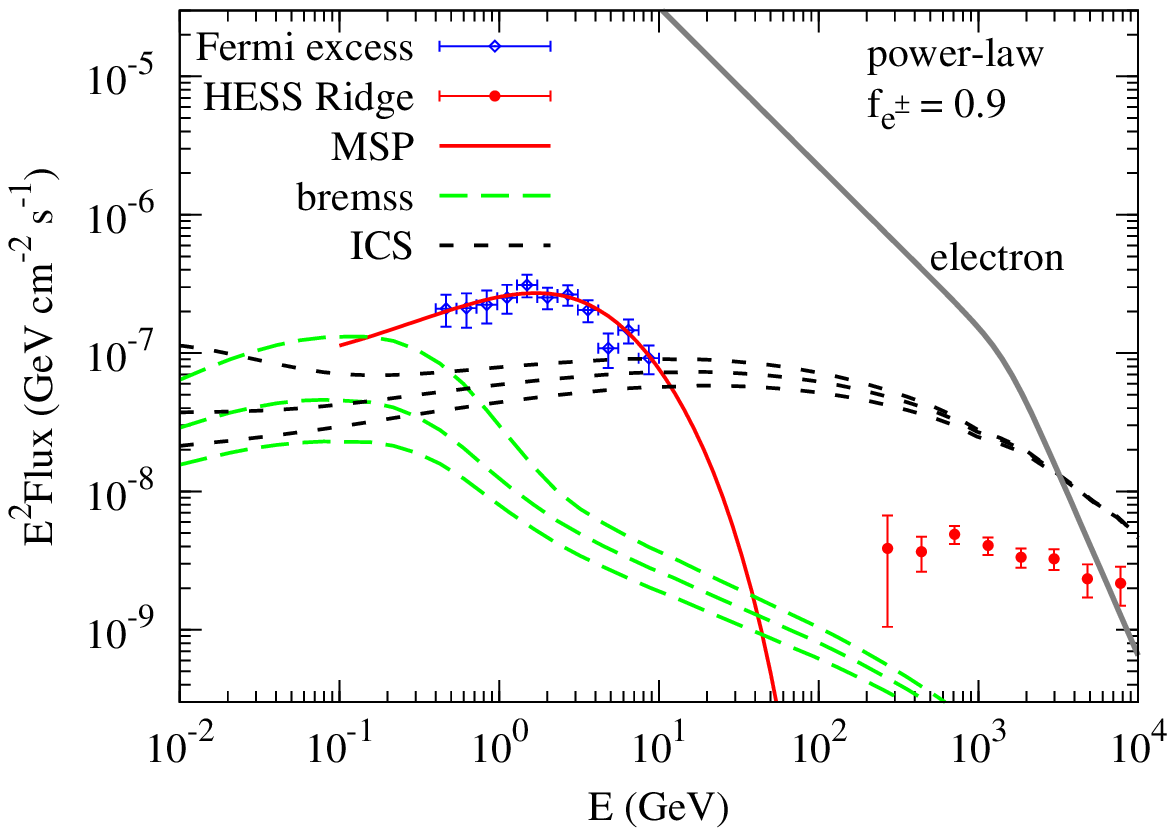}
\includegraphics[width=\columnwidth]{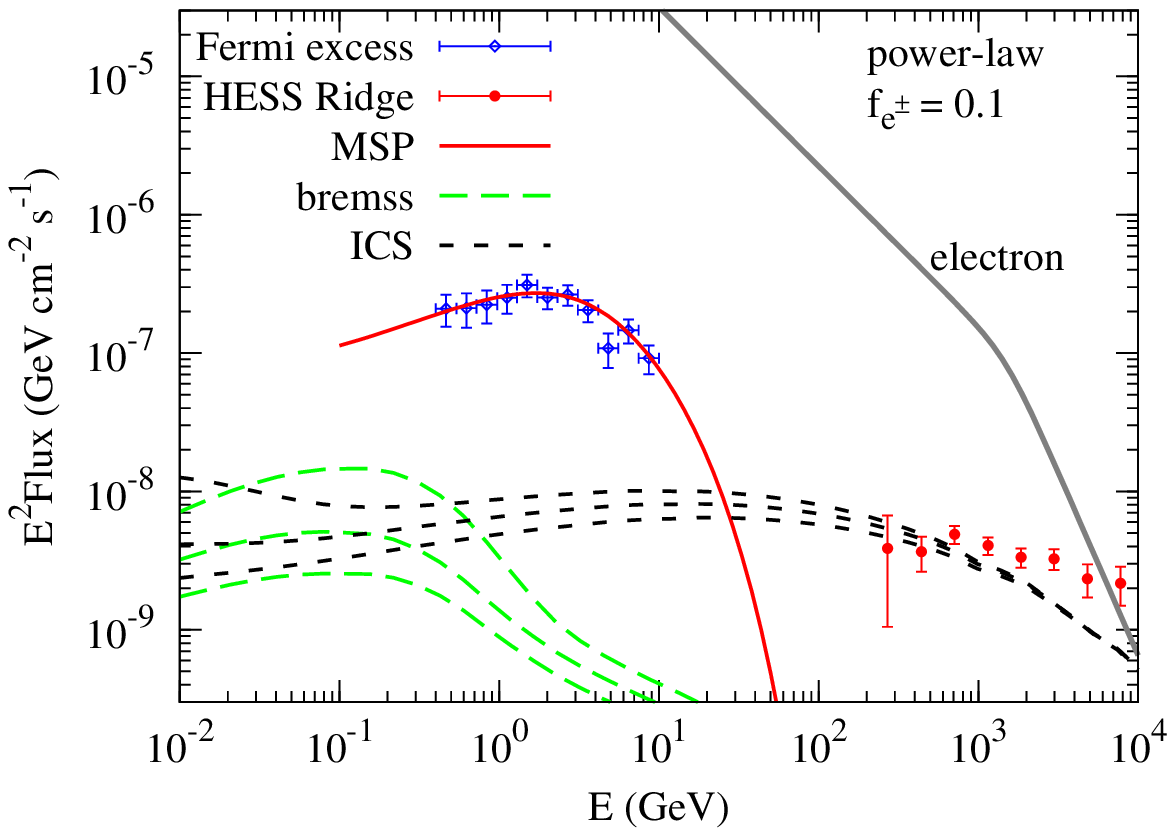}
\caption{Same as Fig. \ref{fig:sed_mono} but for the exponential cutoff
power-law injection of the $e^{\pm}$. The left panel is for $f_{e^{\pm}}=0.9$
and the right panel is for $f_{e^{\pm}}=0.1$.
}
\label{fig:sed_power}
\end{figure*}

The calculated ICS and bremsstrahlung spectra from the propagated
$e^{\pm}$ produced by the MSPs in $7^{\circ}\times7^{\circ}$
region around the GC are shown in Figs. \ref{fig:sed_mono} and
\ref{fig:sed_power}, for the monochromatic injection and the power-law
injection, respectively. The Fermi-LAT data of the $\gamma$-ray excess
\citep{2013PhRvD..88h3521G}, together with the direct $\gamma$-ray
emission expected from a population of MSPs \citep{2014JHEAp...3....1Y}
are also shown. The results show that there is indeed a VHE component, 
mainly from the ICS off the starlight, if the injection energies of the 
$e^{\pm}$ are high enough. For the monochromatic injection case 
with $E_{\rm inj}=20$ GeV, the ICS emission falls in the same energy 
window of the current GC excess. It may contribute partially
to the observed signal. However, the spatial morphology of this ICS
component should be different from the direct $\gamma$-ray emission 
from MSPs (see Fig. \ref{fig:sky1-10}). For $E_{\rm inj}=200$ GeV case, 
if $f_{e^{\pm}}=0.9$ the ICS emission will exceed the Fermi-LAT data above 
several GeV. It may indicate that the $e^{\pm}$ efficiency may not be as 
high as $\sim 1$, or such a high energy tail is missed in the Fermi-LAT 
data analysis because of the different morphology. There is a bump of the 
$\gamma$-ray emission at $10-200$ GeV corresponding to $E_{\rm inj}=200$ 
GeV, and might be detectable by future VHE observations. For the power-law 
injection model, the spectrum of the VHE component is shallower and can 
probably extend to much higher energies.

\begin{figure*}[!htb]
\centering
\includegraphics[width=0.66\columnwidth]{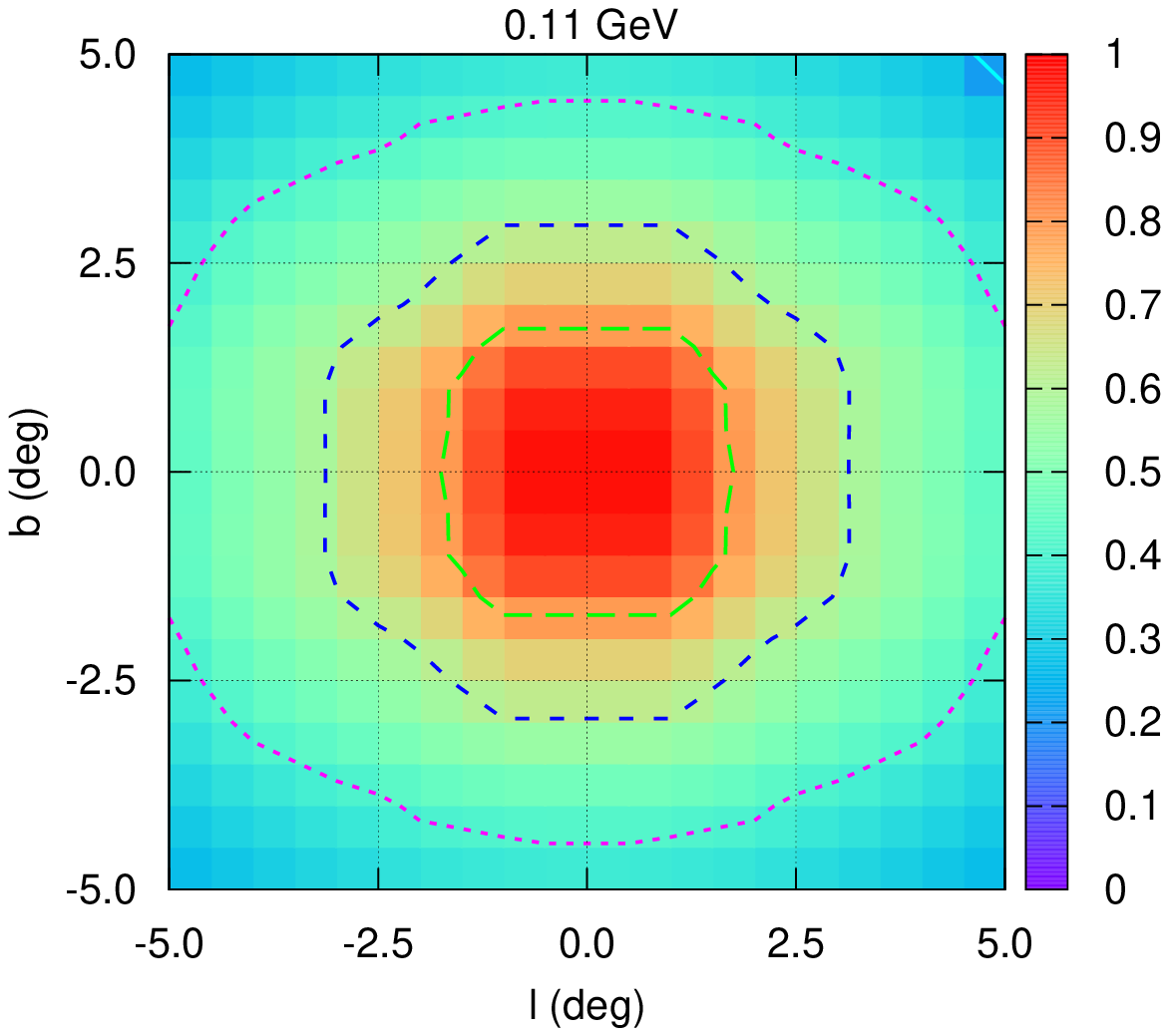}
\includegraphics[width=0.66\columnwidth]{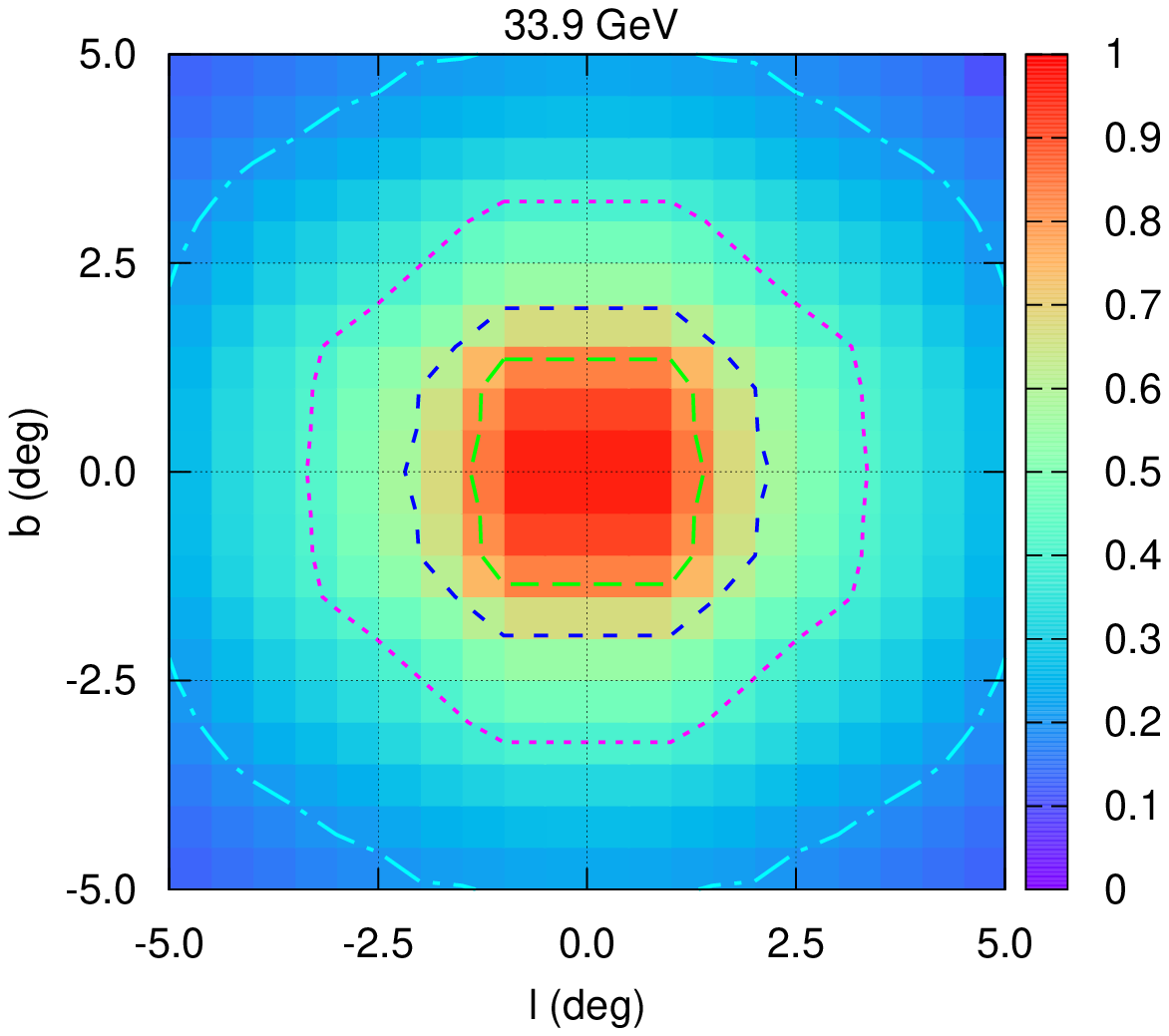}
\includegraphics[width=0.66\columnwidth]{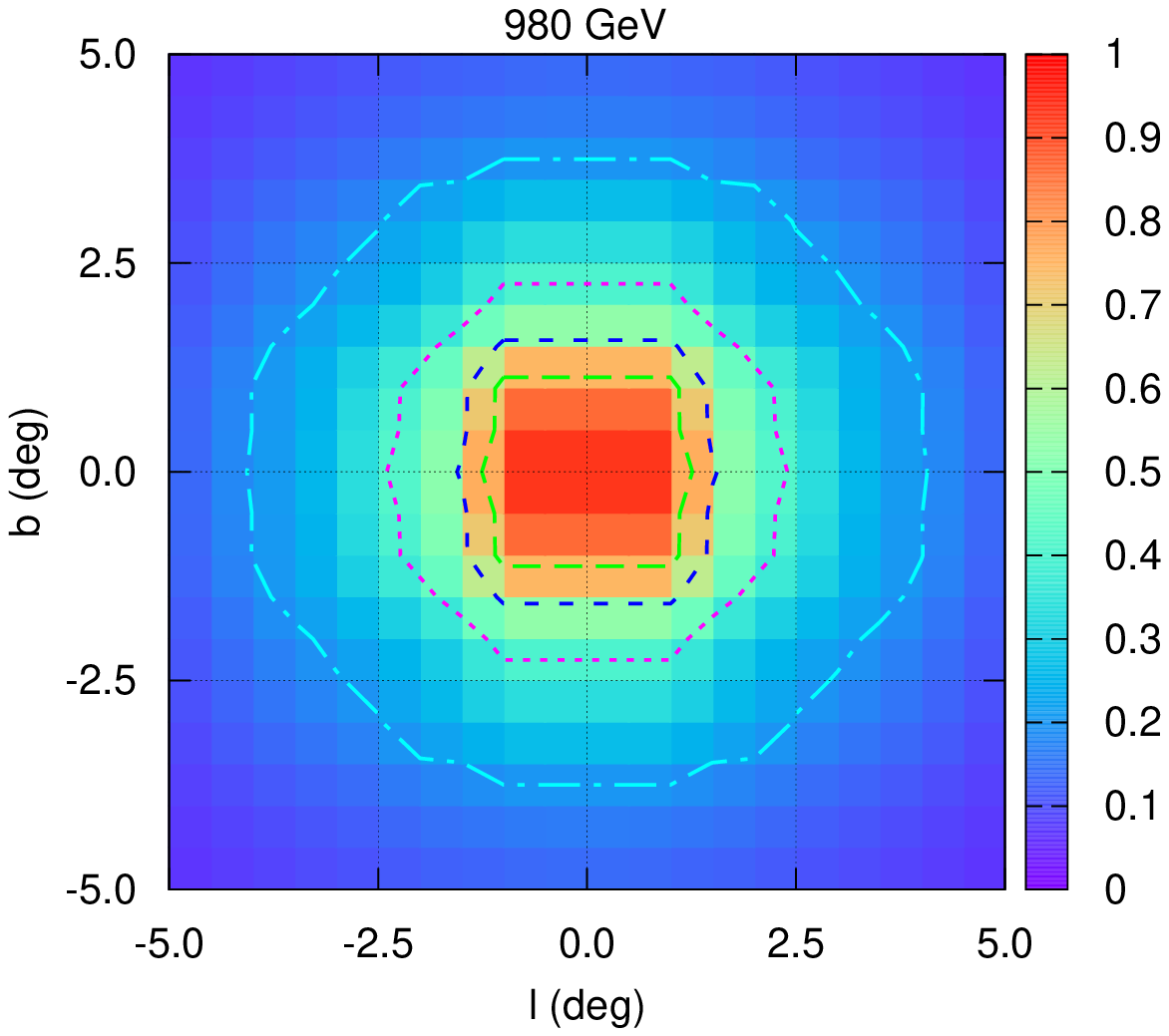}
\includegraphics[width=0.66\columnwidth]{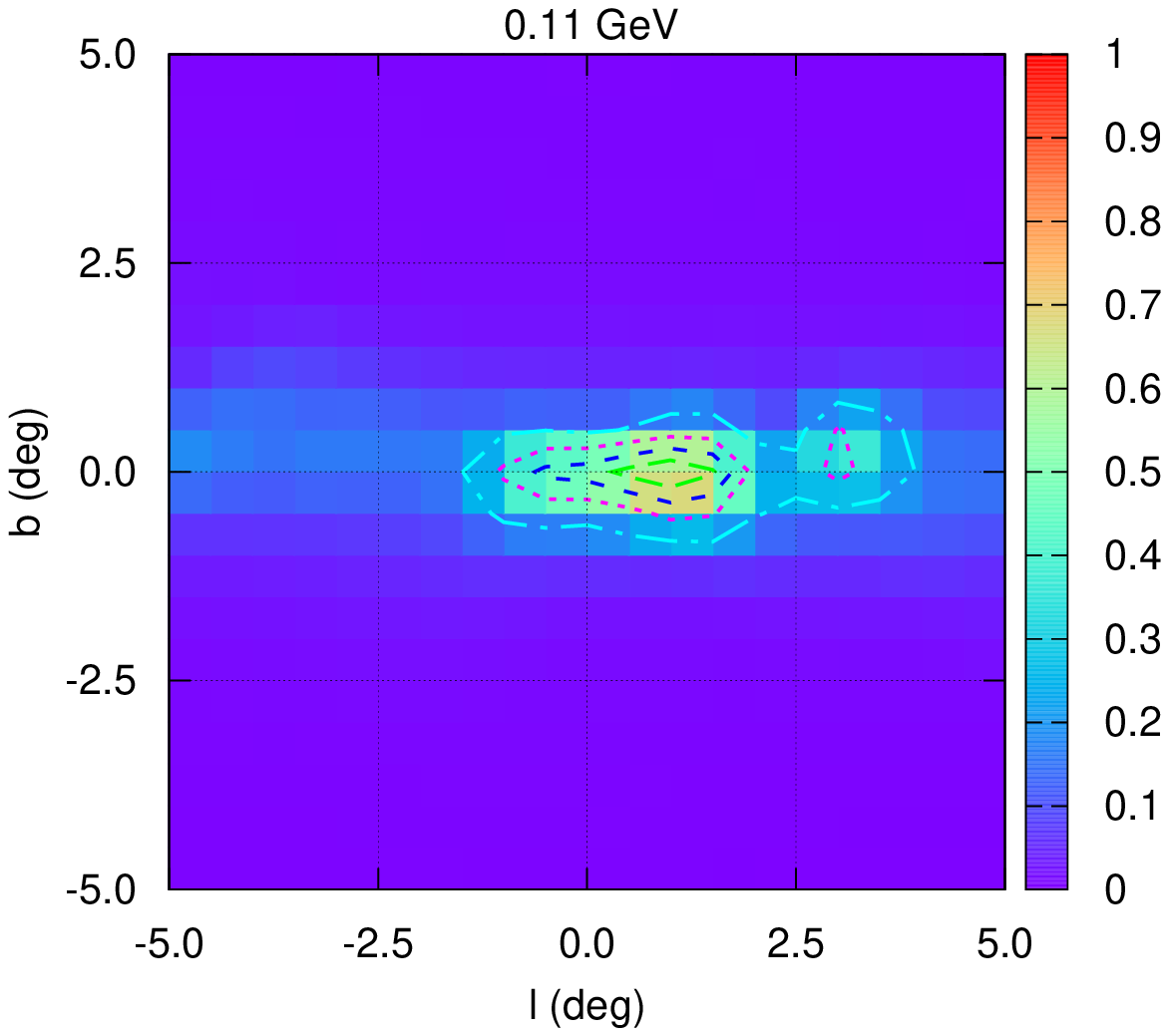}
\includegraphics[width=0.66\columnwidth]{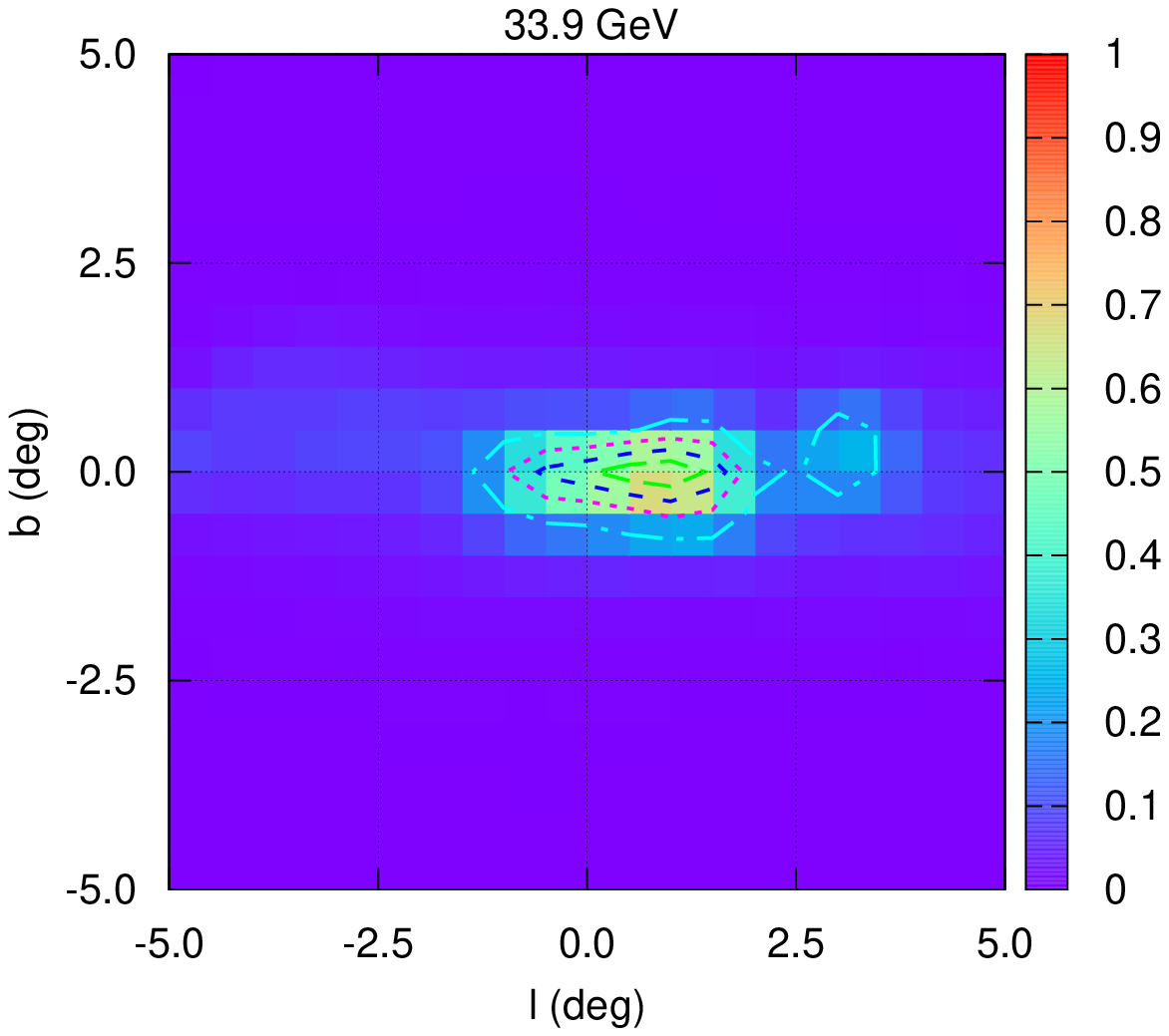}
\includegraphics[width=0.66\columnwidth]{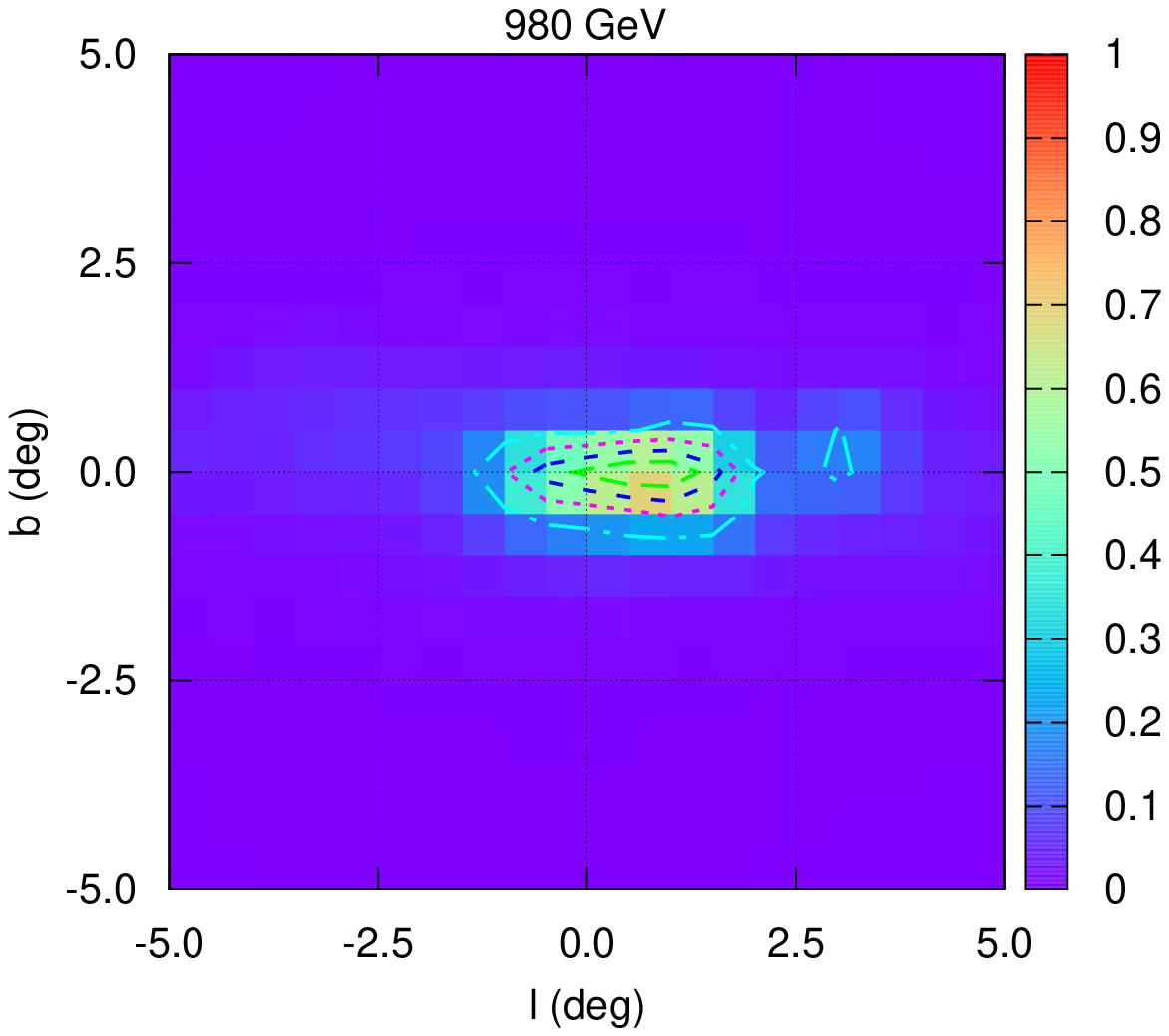}
\caption{Skymaps of the ICS (top) and bremsstrahlung (bottom) emission 
in the central $10^{\circ}\times 10^{\circ}$ region around the GC. 
Contours show the flux levels 0.8, 0.6, 0.4 and 0.2 of the peak value
from inside to outside.
The injection model is the exponential cutoff power-law one in Table
\ref{table:model}, and the propagation parameters are the second group
in Table \ref{table:prop}. For each panel we normalize the maximum flux 
to $1$ to show the relative image. Three different energies, 0.11, 33.9 
and 980 GeV, are shown from left to right. 
}
\label{fig:skymap}
\end{figure*}

In order to have a rough idea of the detectability of this VHE emission, 
we compare it with the HESS detected diffuse ridge emission, integrated
in the region $|l|<0^{\circ}.8$, $|b|<0^{\circ}.3$ in Figs. 
\ref{fig:sed_mono} and \ref{fig:sed_power} \citep{2006Natur.439..695A}. 
The VHE component from the MSPs could be comparable or even brighter 
than the ridge emission, although it is more extended. Given the much 
better sensitivity of the future VHE experiments such as CTA, it should 
be possible to detect such a bright source \citep{2013APh....43....3A}. 

Due to the extension of the emission, the nearly isotropic electron 
background will contaminate the detection. The spectrum of the total 
$e^{\pm}$ based on AMS-02 \citep{2014PhRvL.113l1102A} and HESS 
\citep{2008PhRvL.101z1104A} measurements are also shown by the gray
line in Figs. \ref{fig:sed_mono} and \ref{fig:sed_power}. We can see 
that for the power-law injection scenario with a large value of 
$f_{e^{\pm}}$, the ICS emission may exceed the electron background 
for energies higher than several TeV (the left panel of Fig. 
\ref{fig:sed_power}). However, in general the electron background 
is much higher than the expected $\gamma$-ray signal. Therefore, the
spatial morphology may be necessary to subtract the electron background.

Fig. \ref{fig:skymap} shows the morphology of the ICS (top panels)
and bremsstrahlung (bottom panels) emission, for the power-law injection
scenario. From left to right we show the results at $0.11$, $33.9$
and $980$ GeV, respectively. The ICS emission is almost symmetric
around the GC, and shows a decrease beyond $\sim 2^{\circ}$. The
morphology becomes more concentrated with the increase of energy,
which is expected due to the more significant cooling of higher energy
$e^{\pm}$. The morphological study of the Fermi-LAT $\gamma$-ray
excess shows that the signal is pretty symmetric with the axis ratio
departing from unity less than $20\%$ \citep{2014arXiv1402.6703D}. 
However, the current Fermi-LAT data may not be able to exclude a slightly
asymmetric morphology. The CTA, on the other hand, may determine more
precisely the morphology of this signal if the flux is high. 
We should keep in mind that there are more complexities about the very 
detailed morphology of the ICS emission, caused by the particle
diffusion and the distribution of the background radiation field.
The bremsstrahlung emission is strongly correlated with the
gas distribution, and is more concentrated in the Galactic plane. 
It may be contaminated by the diffuse $\gamma$-ray background from
the Galactic CRs. As we have seen before, the ICS emission will generally
dominate the bremsstrahlung one. Therefore, the nearly symmetric ICS 
component of the VHE emission is an important expectation of the MSP 
model to explain the Fermi-LAT GC excess. The morphology of the ICS 
emission will be crucial to eliminate the electron background.

\begin{figure}[!htb]
\centering
\includegraphics[width=\columnwidth]{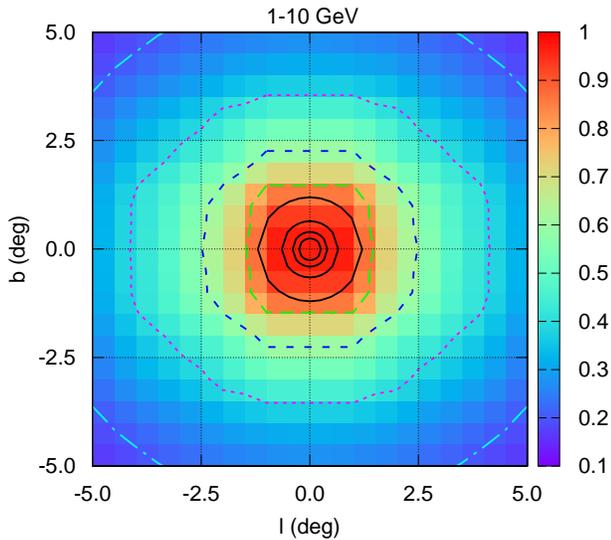}
\caption{Skymap of the ICS emission integrated in $1-10$ GeV range.
Other parameters are the same as Fig. \ref{fig:skymap}. The solid (black)
lines are the contours with flux levels 0.8, 0.6, 0.4 and 0.2 of the 
peak value (inetgrated within $0^{\circ}.1$) from inside to outside, 
as shown in the Fermi-LAT data of the GeV $\gamma$-ray excess 
\citep{2013PhRvD..88h3521G}.
}
\label{fig:sky1-10}
\end{figure}

As a comparison, we show in Fig. \ref{fig:sky1-10} the skymap of 
the ICS emission integrated in $1-10$ GeV range, and the contours of
the observed Fermi-LAT GeV $\gamma$-ray excess \citep[Fig. 4 of][]
{2013PhRvD..88h3521G}. The model parameters to calculate the ICS emission 
are the same as those in Fig. \ref{fig:skymap}, and the flux levels of 
the contours of Fermi-LAT data are 0.8, 0.6, 0.4 and 0.2 from inside 
to outside. The result does show that the direct emission from the MSPs 
is much more concentrated than the ICS emission.

\begin{figure}[!htb]
\centering
\includegraphics[width=\columnwidth]{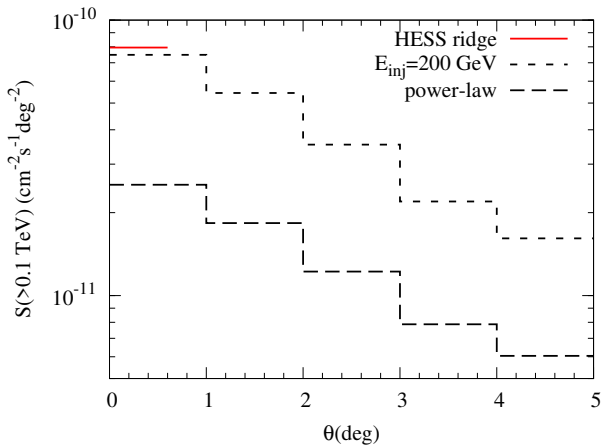}
\caption{Surface brightness distribution of the ICS component from the 
wind $e^{\pm}$ of the MSPs. The second setting of propagation parameters 
in Table \ref{table:prop} is adopted. Here $f_{e^{\pm}}=0.9$ is assumed, 
and for the monochromatic injection scenario we adopt $E_{\rm inj}=200$ 
GeV. The HESS observation of the diffuse emission from the GC ridge is 
also plotted \citep{2006Natur.439..695A}.
}
\label{fig:surface}
\end{figure}

Finally we give the surface brightness distribution of the ICS emission
above $0.1$ TeV in Fig. \ref{fig:surface}. The propagation parameters 
are the second ones in Table \ref{table:prop}. We adopt the efficiency 
$f_{e^{\pm}}=0.9$ for both the monochromatic and the power-law injection 
scenarios, and $E_{\rm inj}=200$ GeV for the monochromatic injection 
scenario. The surface brightness decreases with the increase 
of $\theta$, angle away from the GC. The rough behavior beyond
$1^{\circ}$ is $\sim\theta^{-0.9}$, which is shallower than that of the
Fermi-LAT excess. It is mainly due to the diffusion of the $e^{\pm}$.
The HESS observation of the GC ridge emission, calculated from the 
fitting spectrum $1.73\times10^{-8}(E/{\rm TeV})^{-2.29}$ 
TeV$^{-1}$cm$^{-2}$s$^{-1}$sr$^{-1}$ \citep{2006Natur.439..695A}, 
is shown by the short-dashed line. The ICS emission seems to be fainter 
than the ridge emission. Since the sensitivity of CTA will be better by 
order of magnitude than HESS, it is possible for CTA to identify such a 
VHE $\gamma$-ray component.

\section{Conclusion and discussion}

The recent discovery of the GeV $\gamma$-ray excess from the GC in
Fermi-LAT data has triggered extensive discussions of the DM origin or 
astrophysical origins such as MSPs. It is very essential to investigate 
the possible effects for future observations which may test different 
models. For the MSP scenario, it is expected that the wind $e^{\pm}$, 
with very high Lorentz factor ($\sim10^6$), will carry most of the pulsar 
spin-down energy, and may diffuse and radiate in the vicinity of the GC. 
The ICS and bremsstrahlung emission from those wind $e^{\pm}$ may be 
detectable by the VHE $\gamma$-ray experiments such as CTA. 

We find that there could be a bright VHE component from the wind $e^{\pm}$,
mainly from the ICS off the interstellar radiation field, for a variety
of model parameters. This ICS emission is nearly spherically symmetric
within several degrees of the GC. The spectrum depends on the assumption
of the $e^{\pm}$ injection, and can likely extend to higher than $100$
GeV for reasonable model parameters. The ICS emission has larger total
flux than the Galactic ridge emission observed by HESS 
\citep{2006Natur.439..695A}, although the former is more extended than
the latter, and may be detectable with CTA. The detection of a VHE 
counterpart of the GeV $\gamma$-ray excess will provide a strong support 
of the MSP (or other similar astrophysical) scenario to explain the GeV 
$\gamma$-ray excess compared with the DM annihilation scenario.

The detailed detectability of the VHE source with CTA relies on the detector 
performance and background rejection technique for extended source analysis. 
These are interesting future works. Through a rough comparison with the HESS 
data of the GC ridge and the improvement of the sensitivity of CTA compared 
with HESS, it should be promising for CTA to detect such VHE emission.

We should note that the globular clusters, which are thought to be
rich of MSPs, have been shown to be a class of $\gamma$-ray sources by
Fermi-LAT \citep{2010A&A...524A..75A}. Similar to the scenario discussed
in this work, the VHE $\gamma$-ray emission from the ICS of the wind
$e^{\pm}$ is expected from the globular clusters \citep{2007MNRAS.377..920B,
2009ApJ...696L..52V,2010ApJ...723.1219C}. HESS has discovered the VHE
$\gamma$-ray emission from the direction of globular cluster Terzan 5
\citep{2011A&A...531L..18H}, and set an upper limit for 47 Tucanae
\citep{2009A&A...499..273A}. The detectability of globular clusters with
CTA has been investigated in detail in \citet{2013APh....43..287D}. 
Here we just check the consistency of the model prediction with the current
observations. We take the power-law model as an illustration since the
threshold energies of the HESS observations are relatively high (a few 
hundred GeV). Furthermore we adopt $f_{e^{\pm}}=0.9$ to give a maximum
estimate. Assuming the ratio between the ICS component and the direct 
emission from MSPs for the globular clusters is the same as the GC excess 
discussed in this work, we find that the predicted VHE $\gamma$-ray flux 
is $I(>440\,{\rm GeV})\approx2.9\times10^{-12}$ cm$^{-2}$s$^{-1}$ for 
Terzan 5, and $I(>800\,{\rm GeV})\approx4.4\times10^{-13}$ cm$^{-2}$s$^{-1}$ 
for 47 Tucanae, based on the Fermi-LAT fluxes \citep{2010A&A...524A..75A}. 
As a comparison, the HESS observations give $I(>440\,{\rm GeV})=(1.2\pm0.3)
\times10^{-12}$ cm$^{-2}$s$^{-1}$ for Terzan 5 \citep{2011A&A...531L..18H}, 
and $I(>800\,{\rm GeV})<6.7\times 10^{-13}$ cm$^{-2}$s$^{-1}$ for 47 Tucanae 
\citep{2009A&A...499..273A}. The expectation of the VHE flux for Terzan 5 
seems to be larger by a factor of $\sim2$ than the HESS observation. It
may indicate that $f_{e^{\pm}}$ should not be as high as $\sim1$. However,
we note that there are other uncertainties which also affect the prediction
of the VHE fluxes, such as the magnetic field inside the clusters, the
injection spectrum and diffusion of electrons \citep{2007MNRAS.377..920B}.

Another counterpart to the GeV $\gamma$-ray excess is obviously the 
radio MSPs. The Square Kilometre Array (SKA) will provide a complete census 
of pulsars in the Galaxy \citep{2004NewAR..48.1413C,2009A&A...493.1161S} and 
complement the CTA observations.

{\it Note:} --- When this work is at the final stage, we are aware of a
similar work arXiv:1411.2980 \citep{2014arXiv1411.2980P} which discussed
the ICS emission from the MSPs.

\acknowledgments

We appreciate the anonymous referee for helpful suggestions on this paper.
We thank M. Hayashida, D. Hooper, S. Kisaka, K. Kohri, T. Saito, H. Takami 
and B. Zhang for useful comments. 
This work was initiated during Q.Y.'s stay in KEK as a Short-Term Invited
Fellow. This work is supported by KAKENHI 24000004, 24103006, 26287051 
and 26247042 (K.I.).


\end{document}